%% file: sn_article.tex
\documentclass[
  runningheads  
]{svjour3}
\input{q-dimacs}

\begin{document}
\smartqed


\title{Performance Gains in Quantum SAT Solvers Using ESOP Encoding
}

\titlerunning{Performance Gains in Quantum SAT Solvers}

\author{
  Majd~Assaad \and
  Abhoy~Kole \and
  Rolf~Drechsler
}

\authorrunning{Majd~Assaad et al.}

\institute{
  Majd~Assaad \and Abhoy~Kole \at
  Cyber-Physical Systems, DFKI, Germany \\
  \email{\{majd.assaad,abhoy.kole\}@dfki.de}
  \and
  Rolf~Drechsler \at
  Institute of Computer Science, University of Bremen/DFKI, Germany \\
  \email{drechsler@uni-bremen.de} 
}

\date{Received: XXXX / Accepted: XXXX}

\maketitle

\begin{abstract}
The Boolean Satisfiability (SAT) problem is a canonical NP-complete problem and a
natural candidate for quantum acceleration via search-based algorithms.
In Grover-based quantum SAT solvers, the dominant computational cost stems from
the construction of a reversible oracle that evaluates the Boolean formula,
rendering the choice of SAT encoding crucial for overall quantum resource
efficiency. Although SAT instances are conventionally expressed in Conjunctive
Normal Form (CNF), such encodings typically translate into quantum circuits with
significant qubit overhead and high non-Clifford gate complexity.

In this work, we investigate an Exclusive-Sum-of-Products (ESOP)-based CNF
(e-CNF) representation tailored for quantum SAT solving and analyze its impact
on oracle construction. We derive tighter upper bounds on qubit requirements and
Clifford+$T$ gate counts for Grover-based SAT solvers when e-CNF encodings are
employed in place of standard CNF. In addition, we propose a scalable
transformation from Boolean formulas to e-CNF and present a systematic
procedure for interpreting e-CNF representations as reversible quantum
circuits suitable for oracle implementation.
Experimental evaluation on representative SAT benchmarks
demonstrates that the proposed e-CNF-based approach yields substantial and
consistent reductions in quantum resources, including qubit count, $T$-gate
complexity, and circuit depth, when compared to CNF-based oracle constructions.
These results establish e-CNF as an effective quantum-aware SAT encoding that
significantly improves the practicality of oracle-based quantum SAT solving.

\end{abstract}

\section{Introduction}
\input{sections/introduction}

\section{Background}
\input{sections/background}

\section{SAT Encoding}
\input{sections/encoding}


\section{Experimental Results}

\input{sections/result}
\section{Conclusion}
This work investigated the quantum resource requirements of encoding SAT
instances as reversible quantum circuits, with a particular emphasis on
evaluating the benefits of the proposed e-CNF-based encoding over conventional
CNF-based constructions. Our analysis demonstrates that the choice of Boolean
representation plays a decisive role in determining the qubit overhead and
non-Clifford gate complexity of Grover-based SAT oracles.

Both theoretical bounds and experimental results consistently show that
e-CNF-based clause generation leads to substantial reductions in quantum
resources, especially for equivalence ($\Leftrightarrow$) propositions that
commonly arise in circuit equivalence checking. For instance, encoding a
proposition of the form $p_i \Leftrightarrow (a_j \land b_k)$ using a CNF-based
approach requires three additional ancilla qubits and 99 Clifford+$T$ gates,
whereas the corresponding e-CNF-based implementation achieves the same
functionality using only 17 Clifford+$T$ gates and no additional ancilla
qubits. These improvements are further corroborated by benchmark evaluations,
which demonstrate significant reductions in qubit count, $T$-gate complexity,
CNOT count, and circuit depth across a range of SAT instances.

\rev{
While the proposed e-CNF formulation significantly reduces ancilla and Clif-ford+$T$ gate requirements for Grover-based SAT oracles, the presented results should be interpreted as circuit-level resource optimizations. Current noisy interme-diate-scale quantum devices do not yet provide sufficient fault-tolerant capability for meaningful large-scale comparisons. Nevertheless, the proposed reductions improve the feasibility of future fault-tolerant quantum SAT implementations, where oracle complexity constitutes a major component of the overall Grover search cost.}

Looking forward, future work will focus on developing automated frameworks for
estimating tight resource bounds of e-CNF-based quantum circuit realizations and
extending the benchmarking methodology to larger and more diverse SAT
instances. Additional directions include exploring optimization techniques for
multi-controlled gate synthesis tailored to e-CNF structures and investigating
the applicability of e-CNF encodings to other oracle-based quantum algorithms.
Together, these efforts aim to further enhance the scalability and practical
viability of quantum SAT solvers.

\bibliographystyle{unsrt}
\bibliography{qSAT}

\end{document}

%% file: q-dimacs.tex
\usepackage{comment}
\usepackage{mathtools}

\usepackage{graphicx}%
\usepackage{multirow}%
\usepackage{amsmath,amssymb,amsfonts}%
\usepackage{mathrsfs}%
\usepackage[title]{appendix}%
\usepackage{xcolor}%
\usepackage{textcomp}%
\usepackage{manyfoot}%
\usepackage{booktabs}%
\usepackage{algorithm}%
\usepackage{algorithmicx}%
\usepackage{algpseudocode}%
\usepackage{listings}%
\usepackage{natbib}
\usepackage{braket}
\usepackage{subcaption}

\makeatletter
\let\oldtheequation\theequation
\renewcommand\tagform@[1]{\maketag@@@{\ignorespaces#1\unskip\@@italiccorr}}
\renewcommand\theequation{(\oldtheequation)}
\makeatother

\usepackage{hyperref}

\DeclareRobustCommand{\CCX}{\ensuremath{C^{2}X}}
\DeclareRobustCommand{\CCCX}{\ensuremath{C^{3}X}}

\usepackage{booktabs}
\usepackage{siunitx}
\usepackage{pgfplotstable}
\usepackage{pgfplots}
\usepackage{adjustbox}
\pgfplotsset{compat=1.18}
\usepackage{todonotes}
\usepackage{listings}
\usepackage{xcolor}
\usepackage{courier}
\usepackage{comment}
\newif\ifrevision
\revisionfalse 

\newcommand{\rev}[1]{%
  \ifrevision
    {\color{red!80!black}#1}
  \else
    #1
  \fi
}

\definecolor{dimacspink}{RGB}{200, 0, 200}
\definecolor{dimacsgreen}{RGB}{0, 150, 0}

\lstdefinelanguage{dimacs}{
    morekeywords={p, cnf, qcnf},
    morecomment=[l]{c},
    alsoletter={@,!},
    sensitive=true,
}

\lstdefinestyle{dimacs-style}{
    language=dimacs,
    basicstyle=\ttfamily\footnotesize, 
    keywordstyle=\color{blue}\bfseries,
    commentstyle=\color{black},
    stringstyle=\color{red},
    numbers=left,
    numberstyle=\tiny\color{gray},
    escapeinside={(*@}{@*)},  
    stepnumber=1,
    numbersep=5pt,
    tabsize=4,
    breaklines=true,
    breakatwhitespace=false,
    showstringspaces=false,
    columns=flexible,
    backgroundcolor=\color{white},
    literate={0}{{\textbf{0}}}{1}
             {1}{{\textbf{1}}}{1}
             {-}{{\textcolor{red}{-}}}{1}
             {!}{{\textcolor{orange}{!}}}{1}
             {@}{{\textcolor{purple}{@}}}{1}
             {cnf}{\bfseries{\textcolor{teal}{cnf }}}{1}
             {qcnf}{\bfseries{\textcolor{teal}{qcnf }}}{1}
}

%% file: sections/introduction.tex
The \emph{Boolean Satisfiability} (SAT) problem is a cornerstone of both the
theory and practice of computer science.
It was the first problem proven to be NP-complete by the Cook--Levin theorem
in 1971, establishing that every problem in the complexity class NP can be
reduced to SAT in polynomial time~\cite{10.1145/800157.805047}.
As a consequence, an efficient algorithm for SAT would imply efficient
solutions for all problems in NP.
This fundamental role has made SAT a focal point of research across
computational complexity, algorithm design, and practical problem
solving~\cite{769433,10.1007/3-540-49059-0_14,GOMES200889}.

A SAT instance is defined over a finite set of Boolean variables and a Boolean
formula, with the objective of determining whether there exists an assignment
that satisfies the formula.
In practice, SAT instances are most commonly represented in
\emph{Conjunctive Normal Form} (CNF), where the formula is a conjunction of
clauses, each clause being a disjunction of literals~\cite{10.1145/378239.379017,biere2021preprocessing}.
The outcome of a SAT solver is binary: the instance is either \emph{SAT}, if a
satisfying assignment exists, or \emph{UNSAT}, otherwise.
Although CNF is the dominant representation due to its compatibility with
modern solvers, alternative logical forms such as \emph{Negation Normal Form}
(NNF), \emph{XOR Normal Form} (XNF), and hybrid encodings frequently arise in
domain-specific applications, particularly in cryptography and hardware
verification~\cite{Andraschko2024,10.1145/502090.502091}.
However, scalability and efficient transformation between these normal forms
remain challenging~\cite{PLAISTED1986293,10.1145/3551349.3556938}.

Beyond purely logical encodings, SAT has also been modeled using structural and
diagram-based representations.
These include \emph{Circuit-SAT}~\cite{9643505}, where Boolean formulas are represented
as directed acyclic graphs of logic gates, \emph{Binary Decision Diagrams} (BDDs)~\cite{1676819},
and \emph{And-Inverter Graphs} (AIGs)~\cite{10.1145/1146909.1147048}.
While such representations often provide compact structural insights, they
typically introduce additional overhead when translated into solver-compatible
or reversible forms~\cite{brummayer2006local}.

The importance of efficient SAT solving extends far beyond theoretical
interest.
SAT solvers form the computational backbone of numerous applications,
including electronic design automation, formal verification, software testing,
artificial intelligence, cryptanalysis, bioinformatics, network analysis,
game theory, and combinatorial optimization.
Decades of research have produced highly optimized classical algorithms for
SAT solving, including backtracking-based methods such as
\emph{Davis--Putnam--Logemann--Loveland} (DPLL)~\cite{10.1145/368273.368557},
\emph{Stochastic Local Search} (SLS)~\cite{kautz1996pushing}, and \emph{Conflict-Driven Clause
Learning} (CDCL)~\cite{769433}.
These techniques power state-of-the-art solvers, such as the Z3
\emph{Satisfiability Modulo Theories} (SMT) solver~\cite{10.1007/978-3-540-78800-3_24},
and routinely handle industrial-scale instances with millions of variables
and clauses.
Nevertheless, despite their practical effectiveness, the worst-case
computational complexity of SAT remains exponential, and there exist instances
that remain intractable even for modern solvers~\cite{buss2021proof}.

The intrinsic difficulty of SAT has motivated the exploration of quantum and
quantum-inspired approaches to SAT solving.
Several paradigms have been proposed, including \emph{quantum search} via
amplitude amplification, where SAT is treated as an oracle problem and
Grover's algorithm is used to search the space of assignments~\cite{10.1145/1107523.1107524,10.1145/237814.237866}.
Other approaches include adiabatic quantum computing and quantum annealing,
which encode SAT instances as Ising Hamiltonians or \emph{Quadratic Unconstrained
Binary Optimization} (QUBO) problems~\cite{10.1007/978-3-319-66167-4_9,glover2018tutorial},
as well as hybrid quantum--classical algorithms such as the \emph{Quantum Approximate
Optimization Algorithm} (QAOA)~\cite{Mandl_2024}.
Quantum-inspired classical techniques further seek to leverage insights from
quantum mechanics to improve classical SAT solving~\cite{9605275}.

Among these paradigms, Grover's algorithm~\cite{10.1145/237814.237866} is
particularly attractive due to its proven quadratic speedup over classical
brute-force search.
For SAT problems, where the search space grows exponentially with the number
of variables, this speedup is asymptotically significant.
However, the practical performance of Grover-based SAT solvers is dominated by
the cost of constructing the quantum oracle that evaluates the Boolean
formula.
This cost depends critically on the chosen Boolean encoding, as classical
representations such as CNF can lead to excessive gate counts,
ancilla qubit usage, and circuit depth when mapped to reversible quantum
circuits.

A key challenge in applying Grover’s algorithm to SAT lies in efficiently
encoding the problem as a quantum oracle.
This typically requires translating CNF representations into quantum circuits,
often introducing logical equivalence ($\Leftrightarrow$) relations during the
process.
To mitigate the resulting complexity, prior work~\cite{10.1145/3729229}
proposed the use of \emph{Exclusive-Sum-of-Products} (ESOP)-based CNF representations,
referred to as \emph{e-CNF}, for handling equivalence relations, demonstrating
reductions of up to $60\%$ in clause count.
Although originally applied in the context of hardware equivalence checking,
where CNF clauses arise predominantly from $\Leftrightarrow$ relations, this
approach holds promise for quantum SAT solving more broadly.

Direct ESOP-based transformations of equivalence relations enable more
resource-efficient quantum circuit constructions.
In contrast, applying Tseitin transformations~\cite{tseitin_complexity_1983}
prior to circuit synthesis can significantly inflate quantum resource costs.
Motivated by this observation, the present work investigates the quantum
resource complexity of SAT oracle constructions derived from both CNF and
e-CNF encodings, with a particular focus on Clifford+$T$ gate requirements.
We further propose an automated framework for validating these complexity
bounds and benchmarking e-CNF-based quantum circuits against traditional
CNF-based implementations. Specifically, this work makes the following contributions:
\begin{itemize}
    \item We present a tighter upper-bound analysis for Grover-based SAT solving
    using an e-CNF representation.
    \item We propose a scalable transformation framework for converting SAT
    instances into e-CNF form.
    \item We introduce a method for constructing quantum SAT oracles directly
    from e-CNF representations, reducing gate and qubit overhead.
    \item We provide experimental evaluations demonstrating improved quantum
    resource bounds compared to standard CNF-based encodings.
\end{itemize}

%% file: sections/background.tex
\subsection{SAT Representations and Classical Encodings}


\rev{
Let $X=\{x_1,\dots,x_n\}$ be a set of Boolean variables, and let $F$ denote a Boolean formula over $X$. 
We write $F:\{0,1\}^n\rightarrow\{0,1\}$ for the Boolean function it
denotes and $F(x_1,\dots,x_n)$ for its evaluation under an assignment
$x\in\{0,1\}^n$. Where no confusion arises, we identify a formula with the function
it denotes. We write $F\equiv G$ when two formulas denote the same Boolean function,
and $F\equiv_{\mathrm{SAT}} G$ when they are equisatisfiable.
The Boolean Satisfiability (SAT) problem asks whether there exists an assignment
$\alpha:X\rightarrow\{0,1\}$ such that
\begin{align}
F\big(\alpha(x_1),\dots,\alpha(x_n)\big)=1.
\end{align}
}

\paragraph{CNF and ESOP Representations.}
A Boolean formula is in \emph{Conjunctive Normal Form} (CNF) if
\begin{align}
F = \bigwedge_{j=1}^{m} C_j,
\end{align}
where each clause $C_j=\bigvee_{k=1}^{\ell_j} l_{jk}$ is a disjunction of literals
$l_{jk}\in\{x_i,\neg x_i\}$.
Arbitrary Boolean formulas are commonly converted to CNF using
\emph{Tseitin encodings}, which introduce auxiliary variables while preserving
equisatisfiability~\cite{tseitin_complexity_1983}, at the cost of increased
variable and clause counts.

An alternative representation is the \emph{Exclusive Sum-of-Products} (ESOP) form,
\begin{align}
\rev{F} = \bigoplus_{i=1}^{p}\left(\bigwedge_{j\in S_i} l_{ij}\right),
\end{align}
which is closely related to \emph{Algebraic Normal Form} (ANF) and is well-suited for
functions with parity structure~\cite{122597}.
However, ESOP minimization is NP-hard, and both the number and degree of product
terms may grow exponentially~\cite{Riener2020}.

While CNF is dominant in classical SAT solving due to its compatibility with
resolution and clause learning, \rev{ESOP forms are common in logic synthesis and
reversible computation~\cite{4313212,10.1007/978-3-642-20520-0_16}.}
Neither representation is inherently optimized for quantum circuit
implementations, motivating alternative encodings that reduce quantum resource
overheads.

\subsection{Grover’s Algorithm and Oracle-Based SAT Solving}
\rev{
In quantum computing, a single qubit is represented as a vector in the two-dimensional Hilbert space $\mathbb{C}^2$, with computational basis states $\ket{0}$ and $\ket{1}$ expressed using standard Dirac notation. An $n$-qubit system resides in the tensor product space $(\mathbb{C}^2)^{\otimes n}$, and its state can be expressed as a superposition
\begin{equation}
\ket{\psi} = \sum_{x \in \{0,1\}^n} \alpha_x \ket{x},
\end{equation}
where $\alpha_x \in \mathbb{C}$ and $\sum_x |\alpha_x|^2 = 1$. Quantum operations are modeled as unitary transformations, and measurement projects the state onto computational basis states. In this work, we primarily employ gates from the Clifford+T gate library, including the Hadamard (H), Pauli-X, controlled-NOT (CX), and T gates, together with multi-controlled X ($\mathrm{C}^n\mathrm{X}$) operations. The Clifford+T library is widely adopted for fault-tolerant quantum computing. These gates are composed to construct reversible quantum circuits implementing Boolean functions and oracle operators used in quantum search algorithms.
}

\begin{figure}[t]
    \centering
    \includegraphics[width=0.7\linewidth]{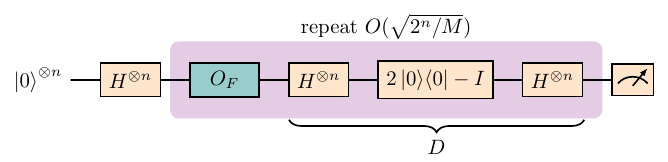}
    \vspace{-0.2cm}
    \caption{Schematic overview of Grover's algorithm for SAT, showing a single
Grover iteration composed of the phase oracle $O_F$ and the diffusion operator
$D$, repeated $O(\sqrt{2^n/M})$ times.}
    \label{fig:GA}
\end{figure}

\rev{Grover’s algorithm enables quantum search over an unstructured space of size
$2^n$ with a quadratic speedup over classical brute-force search.}
Given a Boolean function $F:\{0,1\}^n\rightarrow\{0,1\}$, SAT can be viewed as
searching for $x$ such that $F(x)=1$.
Let
\begin{align}
\mathcal{S}=\{x\in\{0,1\}^n\mid F(x)=1\}, \quad |\mathcal{S}|=M.
\end{align}

The algorithm initializes the uniform superposition
\begin{align}
\ket{\psi_0}=\frac{1}{\sqrt{2^n}}\sum_x\ket{x},
\end{align}
and applies the Grover operator
$G=D\cdot O_F$, where $D=2\ket{\psi_0}\bra{\psi_0}-I$ is the diffusion operator
and $O_F$ is a phase oracle.
After $O(\sqrt{2^n/M})$ iterations as shown in~\autoref{fig:GA}, measurement yields a satisfying assignment
with high probability.

\paragraph{Oracle Construction.}
\rev{A phase oracle is a unitary operator that encodes the value of a Boolean function $F$ into the phase of a quantum state and is defined as
\begin{align}
O_F\ket{x}=e^{i\pi F(x)}\ket{x}.
\end{align}}
\rev{It is typically realized by combining a reversible Boolean oracle $U_F$, which evaluates $F$ onto auxiliary qubit:
\begin{align}
U_F:\ket{x}\ket{0}\mapsto\ket{x}\ket{F(x)}.
\end{align} }
Embedding $F$ into a reversible circuit requires decomposing logical operations
into elementary quantum gates.
CNF encodings induce multi-controlled OR operations, while ESOP encodings
translate to XOR and AND operations with potentially high control degrees.
Consequently, oracle complexity depends strongly on the chosen Boolean
representation.

\paragraph{Overall Complexity.}
\rev{Grover's algorithm requires $O\!(\sqrt{2^n/M})$ iterations of an oracle operator $U_F$ followed by the diffusion operator, where $M$ denotes the number of satisfying assignments and $n$ denotes the number of variables of the Boolean formula $F$. The costs associated with state initialization and the Grover diffusion operator depend primarily on $n$ and are largely independent of the structural complexity of $F$. Therefore, the total circuit complexity becomes

\begin{align}
O\!\left(
\sqrt{\frac{2^n}{M}}
\cdot
(\mathrm{Cost}_F + n)
\right),
\end{align}
where $\mathrm{Cost}_F = |U_F|$ denotes the gate complexity of the oracle circuit $U_F$, measured as the number of Clifford+$T$ gates required to evaluate the Boolean formula $F$ and apply the associated phase flip.

In practical SAT-solving settings, the oracle implementation dominates the overall resource consumption, i.e.,

\begin{align}
\mathrm{Cost}_F \gg O(n),
\end{align}
particularly for large Boolean formulas whose reversible implementations require substantial ancilla management and multi-controlled gate decompositions. Consequently, the overall complexity is dominated by the oracle cost and can be approximated as

\begin{align}
O\!\left(
\sqrt{\frac{2^n}{M}}
\cdot \mathrm{Cost}_F
\right).
\end{align}}

\begin{figure}[t]
    \centering
    \begin{subfigure}{0.45\linewidth}
        \centering
        \includegraphics[width=0.95\linewidth]{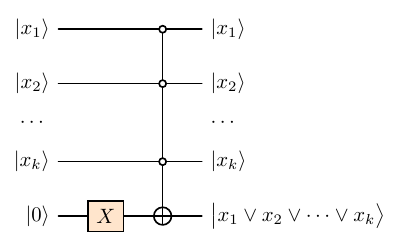}
        \caption{}
        \label{fig:OR}    
    \end{subfigure}
    \hspace{-0.5cm}
    \begin{subfigure}{0.435\linewidth}
        \centering
        \includegraphics[width=0.95\linewidth]{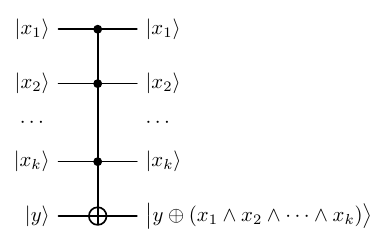}
        \caption{}
        \label{fig:product}
    \end{subfigure}
    \caption{
    Realization of (a) a CNF clause $x_1 \lor x_2 \lor \cdots \lor x_k$ using
input negation and a negative-$k$-controlled Toffoli gate, and
(b) an ESOP product term implementing
$y \oplus (x_1 \land x_2 \land \cdots \land x_k)$
using a positive-$k$-controlled Toffoli gate.}
    \label{fig:cnf-esop}
\end{figure}

\subsection{Quantum Resource Costs and Encoding Challenges}

Although Grover’s algorithm offers quadratic query speedup, its practical
performance is limited by the cost of oracle implementation.
For a SAT instance with $n$ variables, the total number of qubits required is
\begin{align}
Q_{\text{total}} = n + a + 1,
\end{align}
\rev{where $a$ denotes the number of ancilla qubits.}
For CNF encodings derived via Tseitin transformations, $a$ typically scales as
$a=\Theta(m)$, with $m$ clauses, while ESOP encodings require ancillas to
decompose high-degree product terms.

\rev{

The use of ancillary qubits in such constructions reflects a standard trade-off in reversible quantum circuit synthesis. Ancilla-assisted decompositions of multi-controlled Toffoli gates substantially reduce Clifford+T gate complexity and circuit depth compared to ancilla-free realizations~\cite{PhysRevA.93.022311}. The gate counts adopted from~\cite{PhysRevA.93.022311} correspond to previously established Clifford+T decompositions of multi-controlled Toffoli gates and are used here as building blocks for deriving oracle-level resource estimates for CNF- and e-CNF-based constructions.
Although ancilla-free implementations are possible, they typically require more expensive phase-gate decompositions and increased circuit depth~\cite{PhysRevA.106.042602}. Consequently, Tseitin-style constructions are adopted here to obtain resource-efficient fault-tolerant oracle realizations.
}

\begin{figure}[t]
    \centering
    \begin{subfigure}{0.12\linewidth}
        \centering
        \includegraphics[width=0.95\linewidth]{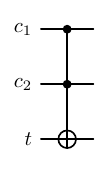}
        \caption{}
        \label{fig:tof}    
    \end{subfigure}
    \hspace{-0.15cm}
    \begin{subfigure}{0.653\linewidth}
        \centering
        \includegraphics[width=0.95\linewidth]{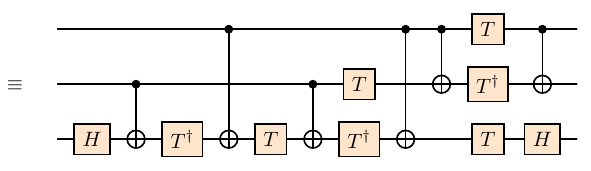}
        \caption{}
        \label{fig:tof-decomp}
    \end{subfigure}
    \caption{(a) A three-qubit Toffoli (CCX) gate and (b) its decomposition into
Clifford+T gates.}
    \label{fig:cliff-t}
\end{figure}

\rev{
Each Grover iteration has cost $\mathrm{Cost}_F + \mathrm{Cost}_D$, where $\mathrm{Cost}_D=O(n)$ is the diffusion
operator cost. The oracle cost $\mathrm{Cost}_F$ dominates whenever $\mathrm{Cost}_F = \Omega(n)$, 
which holds for all SAT encodings considered here. For CNF with $m$ clauses, 
a clause $C_j$ with $\ell_j$ literals induces an $\ell_j$-controlled Toffoli 
gate, whose cost scales linearly in $\ell_j$. The total oracle cost is therefore

\begin{align}
\mathrm{Cost}_F = O\!\left(\sum_{j=1}^{m}\ell_j\right).
\end{align}}
In ESOP encodings, product terms of degree $d$ similarly require multi-controlled
operations with depth growing at least linearly in $d$. 
\autoref{fig:cnf-esop} illustrates Toffoli-network realizations of a CNF clause
and an ESOP product term.

In fault-tolerant settings, non-Clifford gates, particularly $T$ gates, dominate
resource costs.
\autoref{fig:cliff-t} shows the realization of a two-controlled Toffoli (CCX)
gate using a Clifford+$T$ decomposition with $T$-count $7$.
More generally, multi-controlled Toffoli gates incur $T$-counts that scale
linearly with the number of controls, making encodings with wide clauses or
high-degree monomials particularly expensive.

These observations expose a mismatch between classical SAT encodings, which
optimize solver heuristics, and quantum circuit requirements, which favor
shallow circuits with low ancilla and control complexity.
As a result, the effectiveness of Grover-based SAT solving depends critically on
the structural properties of the Boolean encoding, motivating the development
of quantum-aware SAT representations.

%% file: sections/encoding.tex
\subsection{Conventional CNF Representation}

In SAT encoding, the basic logical operators---NOT ($\neg$), OR ($\lor$) and AND ($\land$)---are used to express more complex operations such as implication ($\Rightarrow$), equivalence ($\Leftrightarrow$), and exclusive-OR ($\oplus$) in conjunctive normal form (CNF). This form is particularly suitable because, in practice, many logical formulas emerge from the conjunction of multiple constraints that must hold simultaneously.

\begin{figure}[t]
    \centering
    \includegraphics[width=0.6\linewidth]{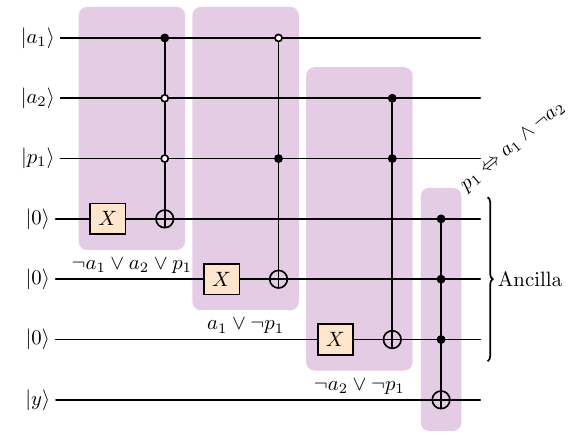}
    \caption{Quantum circuit representation of proposition $p_1  \Leftrightarrow  a_1 \land \neg a_2$ based on conventional CNF interpretation using three ancilla qubits, three Pauli X gates, and a pair of \CCX{} and \CCCX{} gates with some controls of negative polarity.}
    \label{fig:CNF}
\end{figure}

\rev{
The transformation of Boolean formulas into CNF is typically performed using standard logical equivalences, such as De Morgan’s laws to push negations inward (e.g., $\neg(a \lor b) \equiv \neg a \land \neg b$) and distributivity of $\lor$ over $\land$. However, a direct application of these rules may lead to an exponential increase in formula size.

To avoid this blow-up, Tseitin transformations~\cite{tseitin_complexity_1983} introduce auxiliary propositions to represent sub-formulas while preserving equisatisfiability. For example, consider:
\begin{align}\label{eq:formula}
    \phi = (a_1 \land \neg a_2) \lor \neg (a_3 \land a_4) \lor \cdots \lor (a_{n-1} \land a_n).
\end{align}
Introducing fresh variables $p_1, \dots, p_m$ for sub-formulas,
\begin{align}\label{eq:m-propositions}
    p_1 \Leftrightarrow a_1 \land \neg a_2,\; p_2 \Leftrightarrow a_3 \land a_4,\; \dots,\; p_m \Leftrightarrow a_{n-1} \land a_n,
\end{align}
the formula can be rewritten as:
\begin{align}\label{eq:formula2}
    \widehat{\phi} = & (p_1 \lor \neg p_2 \lor \cdots \lor p_m) \nonumber \\
    & \land (p_1 \Leftrightarrow a_1 \land \neg a_2)
    \land \cdots 
    \land (p_m \Leftrightarrow a_{n-1} \land a_n).
\end{align}
While $\phi \not\equiv \widehat{\phi}$, they are equisatisfiable:
\begin{align}\label{eq:eq-sat}
    \phi \equiv_{\text{SAT}} \widehat{\phi}.
\end{align}

Each equivalence constraint is then encoded in CNF. For instance,
\begin{align}\label{eq:AND1}
    p_1 \Leftrightarrow a_1 \land \neg a_2 
    \equiv (\neg a_1 \lor a_2 \lor p_1)\land (a_1 \lor \neg p_1)\land (\neg a_2 \lor \neg p_1).
\end{align}
}

The transformation ensures linear growth in formula size while preserving equisatisfiability, making it a practical tool for SAT solving. 
However, the quantum circuit realization of this transformation remains resource-intensive. Specifically, it requires three additional ancilla qubits per equivalence clause, along with a significant increase in gate complexity, as illustrated in~\autoref{fig:CNF}.

\rev{
According to~\cite{PhysRevA.93.022311}, implementing a three-controlled X ($\text{C}^3\text{X}$) gate requires one ancilla qubit and 33 Clifford+T gates, including 6 Hadamard (H), 15 T, and 12 CNOT (CX) gates. Similarly, a two-controlled X ($\text{C}^2\text{X}$) gate requires 15 Clifford+T gates, consisting of 2 H, 7 T, and 6 CX gates.
Thus, realizing each proposition of the form in~\autoref{eq:AND1} requires two $\text{C}^3\text{X}$ gates, one $\text{C}^2\text{X}$ gate, three logical ancilla qubits for clause evaluation, and three additional Pauli-X gates. The corresponding ancilla and Clifford+T costs are:
\begin{align}\label{eq:ancilla_cnf}
\mathrm{Ancilla}_{\mathrm{CNF}}(p_1 \Leftrightarrow a_1 \land \neg a_2)
= 3 + 2 = 5,
\end{align}
where the additional two ancilla qubits arise from the decomposition of the two $\text{C}^3\text{X}$ gates, and
\begin{align}\label{eq:cost_cnf}
\mathrm{Cost}_{\mathrm{CNF}}(p_1 \Leftrightarrow a_1 \land \neg a_2)
= 2 \times (33 + 15) + 3
= 99.
\end{align}

Similarly, for an OR equivalence constraint:
\begin{align}
p \Leftrightarrow (a \lor b)
\equiv
(p \lor \neg a)\land(p \lor \neg b)\land(\neg p \lor a \lor b),
\end{align}
the resulting quantum realization has the same resource requirements:
\begin{align}
\mathrm{Ancilla}_{\mathrm{CNF}}(p \Leftrightarrow (a \lor b))
= 5,
\end{align}
and
\begin{align}
\mathrm{Cost}_{\mathrm{CNF}}(p \Leftrightarrow (a \lor b))
= 99.
\end{align}

For an XOR equivalence constraint:
\begin{align}
p \Leftrightarrow (a \oplus b)
\equiv &
(\neg p \lor a \lor b)
\land
(\neg p \lor \neg a \lor \neg b)
\land
(p \lor \neg a \lor b)
\land
(p \lor a \lor \neg b),
\end{align}
four clauses must be evaluated and subsequently combined using a $\text{C}^4\text{X}$ gate. The corresponding ancilla and Clifford+T costs become:
\begin{align}
\mathrm{Ancilla}_{\mathrm{CNF}}(p \Leftrightarrow (a \oplus b))
= 4 + 4 + 1
= 9,
\end{align}
where the terms correspond to clause-evaluation ancilla qubits, ancilla qubits required for the four $\text{C}^3\text{X}$ decompositions, and one ancilla qubit for the $\text{C}^4\text{X}$ gate, respectively, and
\begin{align}
\mathrm{Cost}_{\mathrm{CNF}}(p \Leftrightarrow (a \oplus b))
= 4 \times 33 + 55
= 187,
\end{align}
where 55 Clifford+T gates correspond to the realization of the $\text{C}^4\text{X}$ gate.

Consequently, for $m$ equivalence propositions, the resource requirements scale linearly in $m$, with the constant factor depending on the operator. For AND/OR-type equivalences, the CNF-based realization requires $5m$ ancilla qubits and $99m$ Clifford+T gates, whereas XOR-type equivalences require $9m$ ancilla qubits and $187m$ Clifford+T gates.
}

\begin{figure}[t]
    \centering
    \includegraphics[width=0.55\linewidth]{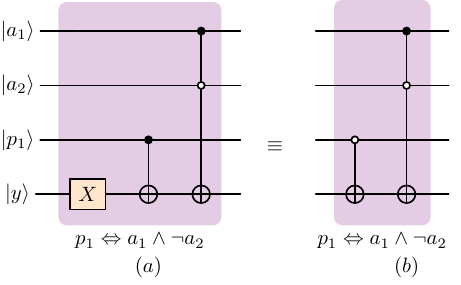}
    \caption{Quantum circuit representations of proposition $p_1  \Leftrightarrow  a_1 \land \neg a_2$ based on e-CNF interpretation using either (a) a Pauli $X$ gate, a $\textit{CX}$ gate, and a \CCX{} gate or (b) a $\textit{CX}$ gate, and a \CCX{} gate with some controls of negative polarity.}
    \label{fig:eCNF}
\end{figure}

\subsection{e-CNF Representation}

\rev{
\paragraph{Definition (e-CNF).}
An \emph{ESOP-based CNF (e-CNF)} formula is a Boolean formula of the form
\begin{equation*}
\tilde{\varphi} = \bigwedge_{k=1}^{m} E_k,
\end{equation*}
where each clause $E_k$ is an Exclusive Sum-of-Products (ESOP) expression, i.e.,
\begin{equation*}
E_k = \bigoplus_{i=1}^{t_k} \left( \bigwedge_{j \in S_{k,i}} l_j \right),
\end{equation*}
with literals $l_j \in \{x_i, \neg x_i, 1\}$, where each $S_{k,i} \subseteq \{1,\dots,n\}$ specifies the variables appearing in the $i$-th product term of clause $E_k$.

Thus, an e-CNF formula is a \emph{conjunction of XORs of conjunctions of literals}. This representation generalizes CNF by replacing disjunctions with exclusive disjunctions (XOR), enabling more compact and expressive encodings of certain logical relations. 
}
Specifically, propositions of the form $p_i\Leftrightarrow \mathcal{F}$ can be equivalently rewritten using XOR as:
\begin{align}
    p_i\Leftrightarrow \mathcal{F} \equiv 1 \oplus p_i \oplus \mathcal{F}.
\end{align}

\rev{For instance, the proposition from~\autoref{eq:AND1} can be represented in e-CNF form as:
\begin{align}\label{eq:AND2}
p_1 \Leftrightarrow a_1 \land \neg a_2
&\equiv
1 \oplus p_1 \oplus (a_1 \land \neg a_2)
\nonumber \\
&\equiv
\neg p_1 \oplus (a_1 \land \neg a_2).
\end{align}


The transformation in~\autoref{eq:AND2} corresponds to the quantum circuit shown in~\autoref{fig:eCNF}, which requires only one $\text{C}^2\text{X}$ gate, one CX gate, one Pauli-X gate, and no additional ancilla qubits. According to~\cite{PhysRevA.93.022311}, the corresponding ancilla and Clifford+T costs are:
\begin{align}\label{eq:ancilla_ecnf}
\mathrm{Ancilla}_{\mathrm{eCNF}}(p_1 \Leftrightarrow a_1 \land \neg a_2)
= 0,
\end{align}
and
\begin{align}\label{eq:cost_ecnf}
\mathrm{Cost}_{\mathrm{eCNF}}(p_1 \Leftrightarrow a_1 \land \neg a_2)
= 15 + 1 + 1
= 17,
\end{align}
where the terms correspond to one $\text{C}^2\text{X}$ gate, one CX gate, and one Pauli-X gate, respectively.

Likewise, the OR equivalence constraint
\begin{align}
p \Leftrightarrow (a \lor b)
&\equiv
p \oplus (\neg a \land \neg b),
\end{align}
requires the same quantum resources:
\begin{align}
\mathrm{Ancilla}_{\mathrm{eCNF}}(p \Leftrightarrow (a \lor b))
= 0,
\end{align}
and
\begin{align}
\mathrm{Cost}_{\mathrm{eCNF}}(p \Leftrightarrow (a \lor b))
= 17.
\end{align}

For XOR equivalence constraints,
\begin{align}
p \Leftrightarrow (a \oplus b)
&\equiv
1 \oplus p \oplus a \oplus b.
\end{align}
the resulting realization requires only CX and Pauli-X gates, yielding:
\begin{align}
\mathrm{Ancilla}_{\mathrm{eCNF}}(p \Leftrightarrow (a \oplus b))
= 0,
\end{align}
and
\begin{align}
\mathrm{Cost}_{\mathrm{eCNF}}(p \Leftrightarrow (a \oplus b))
= 3 + 1
= 4,
\end{align}
where the terms correspond to three CX gates and one Pauli-X gate.

In contrast, the corresponding CNF-based realizations are significantly more resource-intensive. The e-CNF-based approach eliminates the need for multiple ancilla qubits and high-control Toffoli gates introduced through Tseitin-style CNF encodings. The resulting Clifford+T gate reductions achieved using e-CNF over the corresponding CNF realizations are:
\begin{align}
\mathrm{Cost}_{\mathrm{Reduction}}^{\land}
&=
99 - 17
=
82,
\\
\mathrm{Cost}_{\mathrm{Reduction}}^{\lor}
&=
99 - 17
=
82,
\\
\mathrm{Cost}_{\mathrm{Reduction}}^{\oplus}
&=
187 - 4
=
183.
\end{align}

Overall, for $m$ equivalence propositions, the e-CNF-based realization also scales linearly in $m$, while exhibiting substantially smaller constant factors than the corresponding CNF-based construction.}

Additionally, disjunctions ($\lor$) in Boolean formulas can also be represented using ESOP forms, such as:
\rev{
\begin{align}\label{eq:ORn}
    a_1\lor a_2 \lor \cdots \lor a_n \;\equiv\; 1 \oplus \neg a_1 \land \neg a_2 \land \cdots \land \neg a_n.  
\end{align}}
\rev{The resulting formula is a conjunction of ESOP clauses, where each clause is an XOR of conjunctions of literals, consistent with the definition of e-CNF given above.} Using this formulation, the e-CNF representation of~\autoref{eq:formula2} becomes:

\begin{align}\label{eq:formula3}
     \Tilde{\phi} = & (1 \oplus \neg p_1 \land p_2 \land \cdots \land \neg p_m) \land (\neg p_1 \oplus a_1 \land \neg a_2) \nonumber \\
    & \land (\neg p_2   \oplus  a_3 \land  a_4) \land \cdots \land (\neg p_m  \oplus  a_{n-1} \land  a_n).
\end{align}
This representation preserves equisatisfiability, i.e.,
\begin{align}
        \phi &\equiv_{\text{SAT}} \widehat{\phi} \; \equiv_{\text{SAT}} \Tilde{\phi}.
\end{align}
\rev{
Thus, the main advantage of the e-CNF representation arises from the compact encoding of logical equivalence of the form $p \Leftrightarrow \mathcal{G}$. In the absence of such constraints, the quantum resource requirements of CNF- and e-CNF-based realizations are often comparable, since disjunctive clauses can be represented in ESOP form using~\autoref{eq:ORn}. In contrast, when equivalence constraints are present, the CNF construction introduces multiple auxiliary clauses, which lead to additional ancilla qubits and high-control Toffoli gates. For formulas of the form \autoref{eq:formula2}, this yields
\begin{align}
    \mathrm{Ancilla}_{\mathrm{eCNF}}(\widehat{\phi})
    &\leq
    \mathrm{Ancilla}_{\mathrm{CNF}}(\widehat{\phi}), \\
    \mathrm{Cost}_{\mathrm{eCNF}}(\widehat{\phi})
    &\leq
    \mathrm{Cost}_{\mathrm{CNF}}(\widehat{\phi}).
\end{align}
The inequality can be made explicit as follows. According to~\cite{PhysRevA.93.022311}, for $m \geq 2$, the realization of a $\text{C}^m\text{X}$ gate requires $18m-21$ Clifford+$T$
gates, including $4m-6$ H gates, $8m-9$ T gates, and $6m-6$ CX gates. Realizing $m$ propositions of the form in~\autoref{eq:AND2} requires $17m$ Clifford+$T$ gates. The expression $1 \oplus \neg p_1 \land p_2 \land \cdots \land \neg p_m$ additionally requires $18m-20$ Clifford+$T$ gates.

To compute the final output, the conjunction of all $m+1$ ESOP expressions from~\autoref{eq:formula3} requires one additional $\text{C}^m\text{X}$ gate and quantum uncomputation, contributing $18m-21$ and $35m-20$ Clifford+$T$
gates, respectively. Thus, the total gate cost for the e-CNF-based realization is
\begin{align}
    \mathrm{Cost}_{\mathrm{eCNF}}(\phi)
    &= 2\times (35m-20) + 18m-21 \nonumber \\ 
    & = 88m - 61.
\end{align}

In contrast, the conventional CNF-based realization from~\autoref{eq:formula2}
requires
\begin{align}
    \mathrm{Cost}_{\mathrm{CNF}}(\phi)
    &=2\times (117m - 20) + 18m-21 \nonumber \\
    &=252m - 61.
\end{align}

Hence, for this class of logical equivalence encodings, the e-CNF realization reduces the Clifford+$T$ gate count by
\begin{align}
    \mathrm{Cost}_{\mathrm{CNF}}(\phi)
    -
    \mathrm{Cost}_{\mathrm{eCNF}}(\phi)
    =
    164m.
\end{align}
}


\subsection{Quantum Oracle $O_F$ Construction}
Once the CNF or e-CNF representation of the SAT instance is obtained, the quantum
oracle $O_F$ is synthesized using \autoref{alg:synthesize_miter}.
The algorithm takes as input the set of clauses together with their respective
clause types and produces a reversible quantum circuit implementing $O_F$.

The construction begins by initializing an empty oracle circuit consisting of
qubits for all input literals and ancilla qubits corresponding to the total
number of clauses (lines~1--2).
For each clause $C_k$, lines~4--7 describe the gate insertion procedure for
CNF-type clauses, while lines~9--17 specify the corresponding steps for
e-CNF-type clauses.
These steps add the appropriate Toffoli and XOR-based gates required to
reversibly evaluate each clause.

After all clause-specific gates have been appended to the circuit, line~18
constructs the inverse circuit $U^\dagger_F$ by reversing the order of the
previously added gates.
In line~19, a phase-flip operation is applied to mark the computational basis
states corresponding to satisfying assignments of the input CNF or e-CNF
formula.
Finally, line~20 appends the inverse gates to uncompute intermediate values and
restore the ancilla qubits $\mathcal{Y}$ to their initial $\ket{0}$ state.

\begin{algorithm}[t!]
\caption{Synthesize Oracle Circuit}\label{alg:synthesize_miter}
\begin{algorithmic}[1]
\Require Clauses: $\mathcal{C}=\{C_0,\dots,C_{m-1}\}$, Type: $F=\{\text{CNF},\text{e-CNF}\}$
\Ensure Quantum oracle circuit: $O_F$ 
\State $Q \gets \text{getLiterals}(\mathcal{C}) \bigcup$ Ancillae: $\mathcal{Y}=\{y_0,\dots,y_{m-1}\}$ \Comment{Allocate qubits}
\rev{\State $U_F \gets \text{QCircuit}(Q)$}
\For{\textbf{each} clause $C_k \in \mathcal{C}$}
    \If{$F = CNF$}
        \State \rev{$U_{F}\gets U_{F} \cup \Call{X}{y_k}$}
        \State $L \gets \text{getNCLiteralQubits}(C_k)$ \Comment{Get non-complemented literals in $C_k$}
        \State \rev{$U_{F}\gets U_{F} \cup \Call{X}{L} \cup \Call{MCX}{\text{getLiteralQubits}(C_k), y_k} \cup \Call{X}{L}$}
    \Else  \Comment{Synthesize e-CNF}
        \For{\textbf{each} monomial $M \in C_k$}
            \State $L^{'} \gets \text{getCLiteralQubits}(M)$ \Comment{Get complemented literals in $M$}
            \State \rev{$U_{F}\gets U_{F} \cup \Call{X}{L^{'}} \cup \Call{MCX}{\text{getLiteralQubits}(M), y_k} \cup \Call{X}{L^{'}}$}
            \If{\text{isComplemented}($M$)} \State \Call{X}{$y_k$} \EndIf
        \EndFor
    \EndIf
\EndFor
\State \rev{$U^\dagger_{F}\gets \text{getInverse}(U_F)$} \Comment{Get reverse order of all gates present in $U_F$}
\State \rev{$O_{F}\gets U_{F} \cup \Call{H}{y_i}\cup \mathrm{MCX}(\mathcal{Y}-y_i,y_{i}) \cup \Call{H}{y_i}$} \Comment{$\ket{\mathcal{Y}}=e^{i\pi}\ket{\mathcal{Y}}$ if all $y_k=1$}
\State \rev{$O_{F}\gets O_{F} \cup U^\dagger_{F}$}
\State \Return $O_F$
\end{algorithmic}
\end{algorithm}

\begin{figure}[t!]
     \centering
     \begin{subfigure}[b]{0.74\textwidth}
         \centering
          \includegraphics[width=0.95\linewidth]{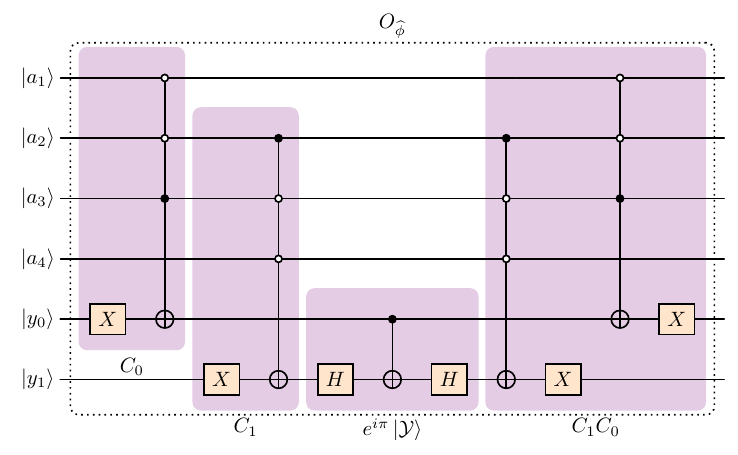}
         \caption{}
         \label{fig:ex-CNF}
     \end{subfigure}
    \begin{subfigure}[b]{0.85\textwidth}
          \centering
          \includegraphics[width=0.95\linewidth]{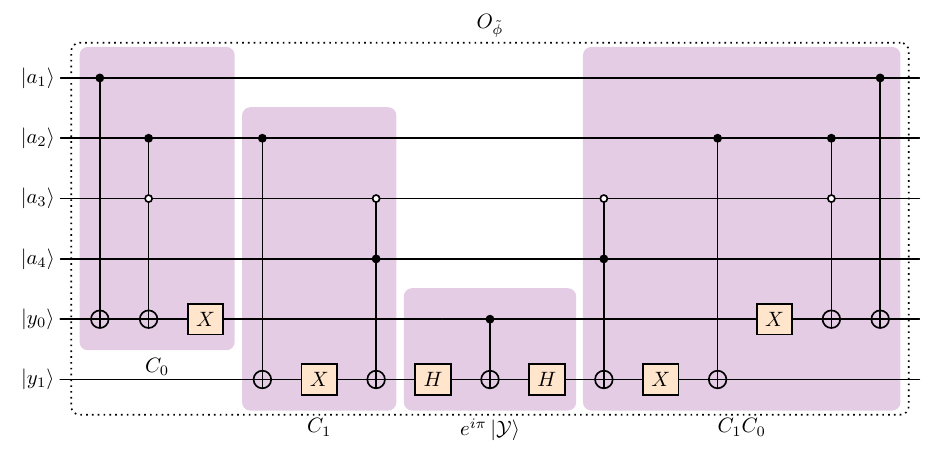}
         \caption{}
          \label{fig:ex-e-CNF}
     \end{subfigure}
  \caption{Quantum oracle circuits synthesized using
\autoref{alg:synthesize_miter}:
(a) oracle $O_{\widehat{\phi}}$ corresponding to the CNF formula
$\widehat{\phi}$ in \autoref{eq:ex-CNF}, and
(b) oracle $O_{\tilde{\phi}}$ corresponding to the e-CNF formula
$\tilde{\phi}$ in \autoref{eq:ex-e-CNF}.} \label{fig:oracle}
\end{figure}

As an illustrative example, consider a CNF formula:
\rev{
\begin{align}\label{eq:ex-CNF}
    \widehat{\phi} &= (a_1 \vee a_2 \vee \neg a_3) \;\land\; (\neg a_2 \vee a_3 \vee a_4),
\end{align}}
and an e-CNF representation:
\rev{\begin{align}\label{eq:ex-e-CNF}
    \tilde{\phi} &= (a_1 \oplus \neg(a_2 \wedge \neg a_3)) \;\land\; (\neg a_2 \oplus (\neg a_3 \wedge a_4)).
\end{align}}
Providing either representation as input to
\autoref{alg:synthesize_miter} yields the corresponding quantum oracle circuits
shown in \autoref{fig:oracle}.

%% file: sections/result.tex
To experimentally evaluate the reduction in quantum resources achieved by the proposed e-CNF-based encoding over the conventional CNF-based approach, we consider a representative set of SAT instances drawn from prior work~\cite{10.1145/3729229}. These benchmarks include:
(i) SAT instances corresponding to equivalence checking of simple Boolean functions composed of 2-input logic gates;
(ii) equivalence checking instances for arithmetic circuits, specifically \emph{carry-ripple} (CR) and \emph{carry-lookahead} (CL) adders; and
(iii) equivalence checking of structurally distinct multiplier designs constructed using combinations of \emph{Unsigned Simple Partial Product} (USP) generators, \emph{Array Multipliers} (AR) with CR adders, and \emph{Wallace Tree} (WT) multipliers with CL adders.

\rev{The SAT benchmarks are generated using the \texttt{qCheck} framework\footnote{\url{https://github.com/qc-agra-cps/qCheck}} from~\cite{10.1145/3729229}.
The proposed oracle synthesis and Clifford+$T$ realization flow are implemented in the \texttt{qSAT} framework\footnote{\url{https://github.com/qc-agra-cps/qSAT}}.
The \texttt{qSAT} tool accepts CNF/e-CNF instances generated from \texttt{qCheck} or equivalent Boolean expressions and synthesizes the corresponding Grover oracle circuits. The generated quantum circuits are exported in OpenQASM~2.0~\cite{cross_open_2017} format.}

\rev{For each SAT instance, quantum oracles are synthesized using the proposed framework in~\autoref{alg:synthesize_miter}. Multi-controlled Toffoli gates arising during oracle construction are decomposed into the Clifford+$T$ gate set using the realization reported in~\cite{PhysRevA.93.022311}, which is explicitly implemented in the proposed framework.} The resulting quantum resource requirements are summarized in~\autoref{tab:result}. The first column lists the benchmark instances, followed by four columns reporting the number of qubits ($\#q$), CNOT gates ($\#CX$), $T$ gates (including $T^\dagger$) ($\#T$), and circuit depth ($\#D$) for oracles constructed using the CNF-based encoding. The next four columns report the corresponding resource metrics for the e-CNF-based encoding.

The final four columns present the relative improvements achieved by the e-CNF-based approach over the CNF-based construction in terms of qubit count, CNOT count, $T$ count, and circuit depth. Overall, the results demonstrate that the proposed e-CNF-based encoding yields substantial resource reductions, achieving up to a $56\%$ decrease in qubits, $81\%$ fewer CNOT gates, $85\%$ reduction in $T$ count, and a $95\%$ reduction in circuit depth compared to the CNF-based approach. 
\pgfplotstableread[
  col sep=semicolon,
  header=false,
  skip first n=2,
  trim cells=true
]{data/CNF_vs_eCNF_results.csv}\CNFeCNFDataTest

\begin{table}[H]
\caption{Comparison of quantum resources (qubits, CNOTs, $T/T^\dagger$ count, and circuit depth)
for CNF- and e-CNF-based SAT oracles across benchmark instances.}\label{tab:result}
\centering
\begingroup
\sisetup{
  group-separator={,},
  output-decimal-marker={.},
  group-minimum-digits=4,
  input-decimal-markers={.},
}
\footnotesize
\setlength{\tabcolsep}{3pt}
\renewcommand{\arraystretch}{1.1}
\begin{adjustbox}{max width=\textwidth}
\pgfplotstabletypeset[
  string replace*={,}{.},
  columns={0,1,2,3,4,5,6,7,8,9,10,11,12},
  columns/0/.style={string type,column type={|>{\raggedright\arraybackslash}m{1.7cm}@{}|},column name=Name},
  columns/1/.style={column type={>{\raggedleft\arraybackslash}m{0.9cm}},column name={\textit{\#q}}},
  columns/2/.style={column type={>{\raggedleft\arraybackslash}m{0.9cm}},column name={\textit{\#CX}}},
  columns/3/.style={column type={>{\raggedleft\arraybackslash}m{0.9cm}},column name={\textit{\#T}}},
  columns/4/.style={column type={>{\raggedleft\arraybackslash}m{0.9cm}|},column name={\textit{\#D}}},
  columns/5/.style={column type={>{\raggedleft\arraybackslash}m{0.8cm}},column name={\textit{\#q}}},
  columns/6/.style={column type={>{\raggedleft\arraybackslash}m{0.8cm}},column name={\textit{\#CX}}},
  columns/7/.style={column type={>{\raggedleft\arraybackslash}m{0.8cm}},column name={\textit{\#T}}},
  columns/8/.style={column type={>{\raggedleft\arraybackslash}m{0.6cm}|},column name={\textit{\#D}}},
  columns/9/.style={column type={>{\raggedleft\arraybackslash}m{0.7cm}},column name={\textit{\#q}}},
  columns/10/.style={column type={>{\raggedleft\arraybackslash}m{0.7cm}},column name={\textit{\#CX}}},
  columns/11/.style={column type={>{\raggedleft\arraybackslash}m{0.7cm}},column name={\textit{\#T}}},
  columns/12/.style={column type={>{\raggedleft\arraybackslash}m{0.7cm}|},column name={\textit{\#D}}},
  every column/.append style={
    fixed,
    /pgf/number format/.cd,
      1000 sep={\text{,}},
      dec sep=.,
    /pgfplots/table/.cd
  },
  columns/1/.append style={precision=0},
  columns/2/.append style={precision=0},
  columns/3/.append style={precision=0},
  columns/4/.append style={precision=0},
  columns/5/.append style={precision=0},
  columns/6/.append style={precision=0},
  columns/7/.append style={precision=0},
  columns/8/.append style={precision=0},
  columns/9/.append style={/pgf/number format/fixed zerofill},
  columns/10/.append style={/pgf/number format/fixed zerofill},
  columns/11/.append style={/pgf/number format/fixed zerofill},
  columns/12/.append style={/pgf/number format/fixed zerofill},
  every head row/.style={
    before row={
      \toprule
      \multicolumn{1}{|c|}{}&\multicolumn{4}{c|}{CNF}&\multicolumn{4}{c|}{e-CNF}&\multicolumn{4}{c|}{Improv. (\%)}\\
      \cmidrule(lr){2-5}\cmidrule(lr){6-9}\cmidrule(lr){10-13}
    },
    after row=\midrule
  },
  every row no 8/.style={after row=\midrule},
  every row no 11/.style={after row=\midrule},
  every last row/.style={after row=\bottomrule}
]{\CNFeCNFDataTest}
\end{adjustbox}
\endgroup
\end{table}